\documentclass[epj]{svjour}
\usepackage{dsfont,amsmath,amssymb,xcolor}
\usepackage{graphicx}
\usepackage{cuted}
\begin{document}
\title{Open-charm Euclidean correlators within heavy-meson EFT interactions}
\author{Gl\`oria Monta\~na\inst{1} \and Olaf Kaczmarek\inst{2,3} \and Laura Tolos\inst{4,5,6,7} \and \`Angels Ramos \inst{1}
}                     
%
%
\institute{Departament de F\'isica Qu\`antica i Astrof\'isica and Institut de Ci\`encies del Cosmos(ICCUB), Facultat de F\'isica, Universitat de Barcelona, Mart\'i i Franqu\`es 1, 08028 Barcelona, Spain
\and Key Laboratory of Quark \& Lepton Physics (MOE) and Institute of Particle Physics,Central China Normal University, 430079 Wuhan, China
\and Fakult\"at f\"ur Physik, Universit\"at Bielefeld, D-33615 Bielefeld, Germany
\and Institut f\"ur Theoretische Physik, Goethe Universit\"at Frankfurt, Max von Laue Strasse 1, 60438 Frankfurt am Main, Germany 
\and Frankfurt Institute for Advanced Studies, Ruth-Moufang-Str.  1, 60438 Frankfurt am Main,Germany
\and Institute of Space Sciences (ICE, CSIC), Campus UAB, Carrer de Can Magrans, 08193 Barcelona, Spain
\and Institut d'Estudis Espacials de Catalunya (IEEC), 08034 Barcelona, Spain
}
\date{Received: date / Revised version: date}
%
\abstract{
The open-charm Euclidean correlators have been computed for the first time using the thermal spectral functions extracted from a finite-temperature self-consistent unitarized approach based on a chiral effective field theory that implements heavy-quark spin symmetry. The inclusion of the full-energy dependent open-charm spectral functions in the calculation of the Euclidean correlators leads to a similar behaviour as the one obtained in lattice QCD for temperatures well below the transition deconfinement temperature. The discrepancies at temperatures close or above the transition deconfinement temperature could indicate that higher-energy states, that are not present in the open-charm spectral functions, become relevant for a quantitative description of the lattice QCD correlators at those temperatures. In fact, we find that the inclusion of a continuum of scattering states improves the comparison at small Euclidean times, whereas differences still arise for large times.  %
} 

\maketitle
\section{Introduction}\label{intro}
Understanding the medium modification of heavy mesons when embedded in a hot and/or dense matter is the subject of ongoing experiments, such as LHC and RHIC, whereas a great effort is also being devoted from the theoretical side (see  \cite{Rapp:2011zz,Tolos:2013gta,Hosaka:2016ypm,Aarts:2016hap} for reviews). Since heavy flavour (charm and beauty) is produced in the early stages of high-energy heavy-ion collisions (HiCs), heavy quarks and their hadronization into heavy mesons turn out to be excellent probes of the properties of the hot and dense medium created during the collisions. In fact, the suppression of quarkonium states, such as the $J/\psi$ meson, in HiCs as compared to proton-proton collisions is widely considered as a signature of the deconfinement of hadronic matter into the quark-gluon plasma (QGP) \cite{Matsui:1986dk}.  This suppression could be also modified due to the in-medium changes of heavy mesons, as described by the comover scattering scenario (see, for example, Refs.~\cite{Capella:2000zp,Cassing:1999es,Vogt:1999cu,Gerschel:1998zi}). 



Lattice QCD is  a powerful tool to study the in-medium modification of heavy mesons through the determination of their spectral properties in matter  (see Ref.~\cite{Rothkopf:2019ipj} for a recent review and references therein). Despite the recent progress in lattice QCD calculations, there are still a few drawbacks that prevent lattice results from being decisive when determining the spectral features of heavy mesons. From lattice QCD one can determine the so-called Euclidean meson correlators and the meson spectral functions are then extracted from them. The reconstruction of the spectral functions from the correlators turns out, however, to be rather complicated \cite{Jarrell:1996rrw}. Furthermore, the simulation of light quarks on the lattice is computationally very demanding and usually larger (unphysical) masses are used. 


Effective field theories in matter offer a complementary strategy to lattice QCD in order to determine the modification of the heavy-meson spectral features in a hot and/or dense medium. Matter below the deconfinement transition temperature consists of hadrons, essentially light mesons, in the low-density high-temperature regime. In this domain, the thermal properties of scalar and vector charm mesons have been recently obtained within a finite-temperature self-consistent unitarized approach based on a chiral effective field theory that implements heavy-quark spin symmetry \cite{Montana:2020lfi,Montana:2020vjg}. Once the spectral features are known, it is then possible to determine the corresponding Euclidean meson correlators and compare to lattice QCD results. In this way, the ill-posed extraction of the spectral function is avoided while testing directly the results from finite-temperature effective field theories against lattice QCD simulations.

In the present paper we determine the Euclidean meson correlators for open charm mesons and compare to the lattice QCD simulations of Ref.~\cite{Kelly:2018hsi}. To the best of our knowledge, this work is the only computation of Euclidean correlators of open-charm mesons. We adapt our calculations of open-charm spectral functions in a pionic bath \cite{Montana:2020lfi,Montana:2020vjg} to the use of the unphysical masses determined in  Ref.~\cite{Kelly:2018hsi}. The paper is organized as follows. In Section \ref{sec:spectral} we introduce the concept of the meson spectral function at finite temperature, while presenting in Sec.~\ref{sec:euclidean} the calculation of the Euclidean correlators and spectral functions in lattice QCD. In Sec.~\ref{sec:spectral-effective} we summarize our calculation for the open-charm spectral function within the effective field theory employing the meson unphysical masses reported in Ref.~\cite{Kelly:2018hsi}, whereas in Sec.~\ref{sec:euclidean-effective} we present our results for the Euclidean correlators and compare them with those from lattice QCD. Finally, in Sec.~\ref{sec:conc} we give our conclusions and future outlook.


\section{Spectral functions of mesons at finite temperature}
\label{sec:spectral}

\begin{figure}[htbp!]
\centering
\includegraphics[scale=0.6]{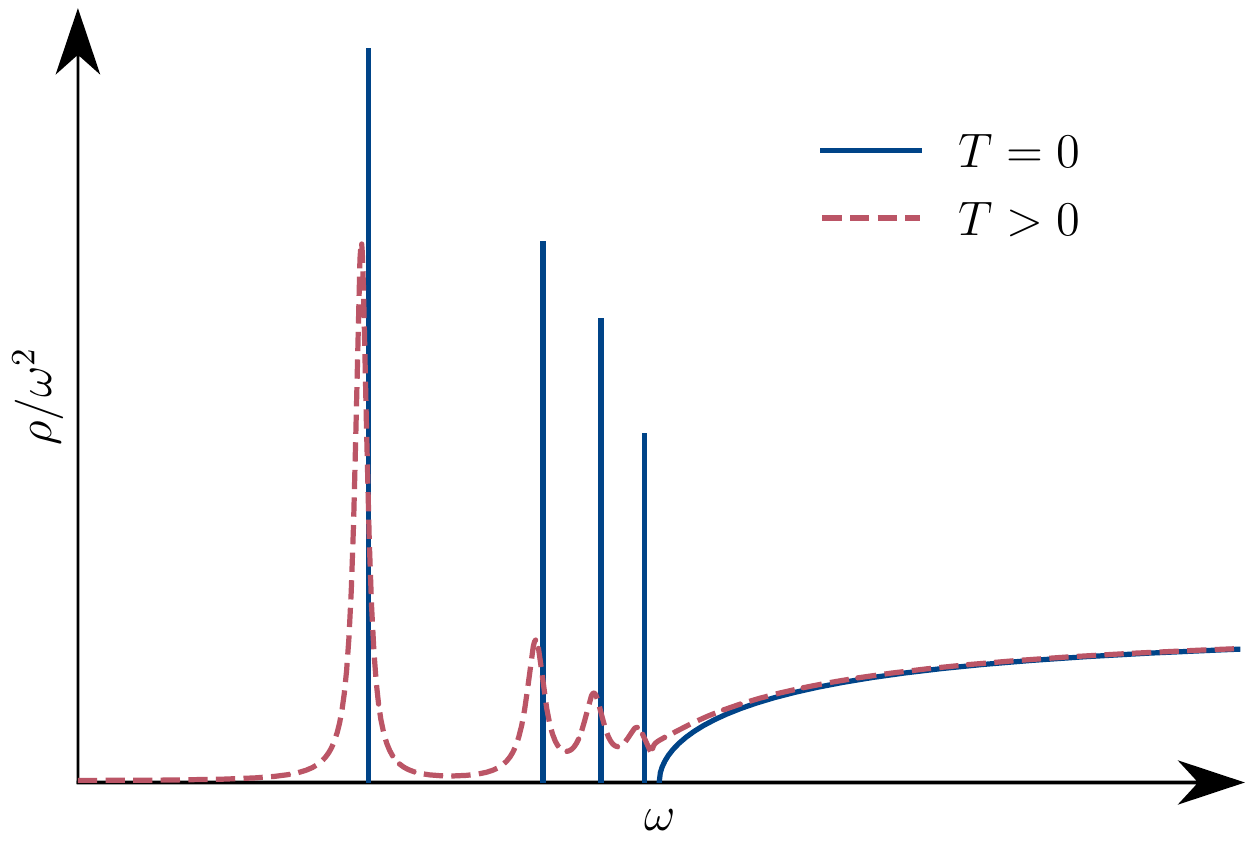}
  \caption{Schematic picture of the meson spectral function at $T=0$ (blue solid line) and at finite temperature (red dashed line).}
  \label{fig:sketch}
\end{figure}

The meson spectral function at finite temperature contains information not only on the mass and width of the ground state but also the masses and widths of the possible excited bound states as well as the continuum of scattering states. A schematic picture is shown in Fig.~\ref{fig:sketch}. At $T=0$ the spectral function results from the contribution of different delta functions corresponding to the ground state of mass $m$ and the bound excited states, and a continuum distribution starting at $2m$ for 2-particle states. At finite temperature, one expects the masses to be modified as well as a broadening of 1-particle states to take place.

There exist a few theoretical approaches that can be used to determine the features of the meson spectral function, but none of them is yet conclusive in the full range of energies and temperatures available in the experiments. Perturbative QCD can be only applied at very large energies and/or temperatures \cite{Karsch:2000gi,Aarts:2005hg,Mocsy:2007yj}. Using lattice QCD, the meson correlator can be calculated from first principles for any energy and temperature (a priori), but the spectral function needs to be reconstructed from the lattice data, which is a non-trivial task \cite{Jarrell:1996rrw,Rothkopf:2019ipj}. The AdS/CFT duality has also been used to describe some features of the spectral function, but a clear correspondence with QCD allowing quantitative studies is still missing \cite{Erdmenger:2007cm,CasalderreySolana:2011us}. Hadronic models based on effective theories at finite temperature are an additional tool to learn about the spectral function below the deconfining temperature \cite{Rapp:2011zz,Tolos:2013gta,Hosaka:2016ypm,Aarts:2016hap}. Therefore, only the interplay between these techniques may shed light on this issue.

\section{Euclidean correlators and spectral functions in lattice QCD}
\label{sec:euclidean}

In lattice QCD the Euclidean space-time is discretized on a 4D grid or lattice of size $N_\sigma^3\times N_\tau$, with lattice spacing $a$. In the limit of vanishing $a$, the continuum theory is recovered. The discretization of space-time introduces a UV cutoff scale, which regularizes the theory in a natural way by restricting the highest momentum to $\Lambda<\frac{\pi}{a}$, and allows one to evaluate the path integrals with Monte Carlo methods using importance sampling.

The fundamental degrees of freedom of lattice QCD are quark fields $\bar\psi_x$, $\psi_x$, that live on the sites of the lattice (labeled with $x\equiv(t,\vec{x})$), and the gluons that reside on the links and are represented by the gauge links $U_{x,\mu}$. 

The primary tools in lattice QCD calculations are Euclidean correlators of some operators $\widehat{O}$, described by the path integral over all degrees of freedom:
\begin{align}\nonumber
 \langle \widehat{O}_1(\tau,\vec{x})&\widehat{O}_2(0,\vec{0})\rangle=\frac{1}{\mathcal{Z}}\int\mathcal{D}[U]\mathcal{D}[\bar{\psi},\psi]  \\ &\times
 \widehat{O}_2[U,\bar\psi,\psi]\widehat{O}_1[U,\bar\psi,\psi]
  e^{-S_F[U,\bar\psi,\psi]-S_G[U]} ,
\end{align}
with $ \mathcal{Z}$ being the partition function
\begin{equation}
 \mathcal{Z}=\int\mathcal{D}[U]\mathcal{D}[\bar{\psi},\psi]e^{-S_F[U,\bar\psi,\psi]-S_G[U]} .
\end{equation}
Each configuration of fields is weighted by the Boltzmann factor $e^{-S}$, where $S_F[U,\bar\psi,\psi]$ is the fermion part and $S_G[U]$ the gluon part of the discretized QCD action. Simulations with dynamical quarks turn out to be very resource demanding and one sometimes uses the quenched appproximation that neglects the fermion action.

For a meson with quantum numbers $H$, the meson (quark anti-quark pair) operator to consider is $J_H(\tau,\vec{x})=\bar{\psi}_f(\tau,\vec{x}),\Gamma_H\psi_f(\tau,\vec{x})$, where $\Gamma_H=\mathds{1},\gamma_5,\gamma_\mu,\gamma_5\gamma_\mu$ correspond to the scalar, pseudoscalar, vector and axial vector channels, respectively, and $f$ refers to the flavour of the valence quark. Carrying out the explicit integration over fermion fields, the propagation of a meson from time $t=0$ to $\tau$ is given by the time correlator
\begin{align}\nonumber
 \langle J(\tau,\vec{x})&J(0,\vec{0})\rangle=-\frac{1}{\mathcal{Z}}\int\mathcal{D}[U]e^{-S_G[U]}{\rm det\,}[M] \\ &\times
 {\rm Tr\,}[\Gamma_HM^{-1}(0,\vec{0};\tau,\vec{x})\Gamma_HM^{-1}(\tau,\vec{x};0,\vec{0})],
\end{align}
with
\begin{equation}
 \mathcal{Z}=\int\mathcal{D}[U]e^{-S_G[U]}{\rm det\,}[M].
\end{equation}
The inverse of the fermion matrix $M^{-1}$ represents the fermion propagator. The effects of the sea quarks are contained in ${\rm det\,}M$. Clover fermions, for which $M$ contains the Wilson and clover terms in addition to the na\"ive discretization of the Dirac operator, are often used. 

In the case of meson correlators at finite temperature one has to consider that the temperature of the system on the lattice is related to the temporal extent through $T=1/(aN_\tau)$. The Euclidean temporal correlator in momentum space
\begin{equation}
 G_E(\tau,\vec{p};T)\equiv\int d^3\vec{x}e^{-i\vec{p}\cdot\vec{x}}\langle J(\tau,\vec{x})J(0,\vec{0})\rangle
\end{equation}
is related to the spectral function $\rho(\omega,\vec{p};T)$ through the convolution with a known kernel $K(\tau,\omega;T)$:
\begin{equation}\label{eq:corr_to_sfunc}
 G_E(\tau,\vec{p};T)=\int_0^\infty d\omega K(\tau,\omega;T)\rho(\omega,\vec{p};T),
\end{equation}
with
\begin{align}\nonumber
 K(\tau,\omega;T)&=\frac{\cosh[\omega(\tau-\frac{1}{2T})]}{\sinh(\frac{\omega}{2T})}=\\ &=e^{\omega\tau}f(\omega,T)+e^{-\omega\tau}[1+f(\omega,T)],
\end{align}
where $f(\omega,T)=[e^{\omega/T}-1]^{-1}$ is the Bose-Einstein statistical factor.
In the following we omit the dependence on the momentum $\vec{p}$, as we focus on spectral functions with $\vec{p}=\vec{0}$ for simplicity, and make the identifications $G_E(\tau;T)\equiv G_E(\tau,\vec{p}=\vec{0};T)$ and $\rho(\omega;T)\equiv\rho(\omega,\vec{p}=\vec{0};T)$.

In lattice QCD simulations values of the Euclidean correlator are obtained for a set of points in Euclidean time, $\tau=\tau_i$, i.e. $\{\tau_i,G_E(\tau_i;T)\}$ for $i=1,N_\tau$ and $\tau_i\in[0,1/T]$. In addition, the lattice data $G_E(\tau_i;T)$ have a statistical error due to the fact that only a finite number of gauge configurations can be generated in a Monte Carlo simulation. The inversion of Eq.~(\ref{eq:corr_to_sfunc}) to extract a continuous spectral function $\rho(\omega;T)$ from such limited number of data points with noise is an ill-posed problem. Two methods are usually employed to try to circumvent this problem, both taking specific assumptions on the shape of the spectral function: i) Bayesian methods like the maximum entropy method (MEM) or stochastic reconstruction methods 
\cite{Ding:2017std}, which perform the kernel inversion by statistically inferring the most probable spectral function; and ii) fitting the lattice data with suitable {\it Ans\"atze} for the spectral function, incorporating bound states, a continuum or perturbative input \cite{Burnier:2017bod}. The fact that a priori assumptions about the spectral function are needed makes the determination of their shape and details at finite temperature very challenging, as we do not have much prior information on them.

There are other error sources that make the results obtained on the lattice differ from the desired physical ones. In addition to the mentioned statistical errors, there are effects tied to the finite lattice spacing $a$, volume effects due to the finite lattice volume, and the large quark masses used in the Monte Carlo calculations. These errors can be minimized by extrapolating to the continuum, taking infinite volume limits and physical mass limits, but they are usually tedious and  not always performed in lattice data analysis.

By inspecting Eq.~(\ref{eq:corr_to_sfunc}), one can see that the temperature dependence of the correlators does not only depend on that of the spectral function but the integration kernel also carries an inherent dependence on the temperature. When directly comparing correlation functions at different temperatures, one may want to discern the differences due to the modification of the spectral function with temperature alone. In order to do so,  it is useful to define the so-called reconstructed correlator at a reference temperature $T_r$,
\begin{equation}\label{eq:corr_reconstructed}
 G_E^r(\tau;T,T_r)=\int_0^\infty d\omega K(\tau,\omega;T)\rho(\omega;T_r).
\end{equation}
The integration kernel is the same as that of $G_E(\tau;T)$ and therefore any difference when comparing $G_E(\tau;T)$ and $G_E^r(\tau;T,T_r)$ arises from differences in the spectral functions at $T$ and $T_r$. The value of $T_r$ is usually chosen to correspond to a temperature at which the shape of the spectral function is better known, so one can reliably trust the spectral function obtained from the lattice correlator. Thus,  the lowest temperature available is usually chosen.

In finite temperature studies of heavy quarks on the lattice, sometimes anisotropic lattices are used, on which the spacing in the temporal direction is smaller than that in the spatial directions (with an anisotropy parameter $\xi=a_s/a_\tau>1$) in order to have a fine enough time discretization, although cut-off effects in spectral functions are determined by the spatial lattice spacing. 

Now, regarding the lattice setup we will use in this paper, we consider the setup of Ref.~\cite{Kelly:2018hsi}, where $a_s=0.123\rm~fm$ and $a_\tau^{-1}=5.63\rm~GeV$, with $\xi=3.5$. The ensembles employed contain dynamical light and strange quarks, with unphysical masses for the two mass-degenerate light quarks and roughly physical values for the strange and charm quarks. The resulting masses of the light and charm mesons on the lattice are the following: $m_\pi=384$ MeV, $m_K=546$ MeV, $m_\eta=589$ MeV, $m_D=1880$ MeV, $m_{D_s}=1943$ MeV. The pseudocritical temperature determined for this configuration is $T_c=185$ MeV and the correlators have been calculated for temperatures $T=(44,\,141,\,156,\,176,\,201,\,235,\,281,\,352)\rm$ MeV. For further details, we refer the reader to Ref.~\cite{Kelly:2018hsi} and references therein.

\section{Spectral functions within an effective theory at finite temperature}
\label{sec:spectral-effective}

In Refs.~\cite{Montana:2020lfi,Montana:2020vjg} we have obtained the spectral functions of the $D$ and $D_s$ mesons in a pionic bath at finite temperature. We use an effective Lagrangian describing the interactions of open heavy-flavour mesons ($D$ and $D_s$) with the light pseudoscalar mesons ($\pi$, $K$, $\bar{K}$ and $\eta$) that is based on chiral and heavy-quark spin-flavour symmetries. The scattering amplitudes are unitarized within the Bethe-Salpeter approach, i.e. solving the matrix equation 
\begin{equation}\label{eq:BS}
T_{ij}=V_{ij}+V_{ik}G_{_{D\Phi},k}T_{kj}
\end{equation}
in a full-coupled-channels basis, where $T$, $V$ and $G_{_{D\Phi}}$ are the unitarized amplitude, the interaction potential and the meson-meson two-body propagator, respectively. The subindices $i,j,k$ denote the incoming and outcoming charm me\-son($D$) - light meson($\Phi$) channels. The effects of a me\-dium at finite temperature are introduced in this model within the imaginary time formalism (ITF) by dressing the mesons in the loop function with the spectral functions as:
\begin{align}\label{eq:loop}
 G_{_{D\Phi}}&(E,\vec{p};T)=\int\frac{d^3q}{(2\pi)^3}\int d\omega\int d\omega' \\ \nonumber &\frac{S_D(\omega,\vec{q};T)S_\Phi(\omega',\vec{p}-\vec{q};T)}{E-\omega-\omega'+i\varepsilon}[1+f(\omega,T)+f(\omega',T)],
\end{align}
where the momentum integration is regularized with a cut-off. We note that here and in Refs.~\cite{Montana:2020lfi,Montana:2020vjg} we have neglected the thermal modification of the pions since the change with temperature is expected to be small \cite{Montana:2020vjg}.

The spectral function of the charm meson is defined in terms of its propagator in a hot medium, $\mathcal{D}_D$,  as
\begin{align}\label{eq:Sfunc}
 S_D(\omega,&\vec{q};T)=-\frac{1}{\pi}{\rm Im\,}\mathcal{D}_D(\omega,\vec{q};T) \\ \nonumber =&-\frac{1}{\pi}{\rm Im\,}\Big(\frac{1}{\omega^2-\vec{q}^2-M_D^2-\Pi_D(\omega,\vec{q};T)}\Big),
\end{align}
where the self-energy, $\Pi_D$, is obtained from closing the pion line in the $T_{D\pi\rightarrow D\pi}$ matrix element of the unitarized amplitude. In the ITF it is given by
\begin{align}\nonumber\label{eq:selfenergy}
 \Pi_D(\omega,\vec{q};T)=&\int\frac{d^3q'}{(2\pi)^3}\int dE\frac{\omega}{\omega_\pi}\frac{f(E,T)-f(\omega_\pi,T)}{\omega^2-(\omega_\pi-E)^2+i\varepsilon} \\  \times &\Big(-\frac{1}{\pi}\Big){\rm Im\,}T_{D\pi\rightarrow D\pi}(E,\vec{p};T),
\end{align}
where $\vec{q'}=\vec{p}-\vec{q}$.
The set of Eqs.~(\ref{eq:BS}) to (\ref{eq:selfenergy}) needs to be solved iteratively until convergence is achieved to ensure the self-consistency of the results. In this hadronic effective model the masses of the mesons in vacuum are input parameters and, therefore, can be tuned to those used in lattice QCD for a straightforward comparison of both approaches. 

\begin{figure}[htbp!]
\centering
\includegraphics[scale=0.7]{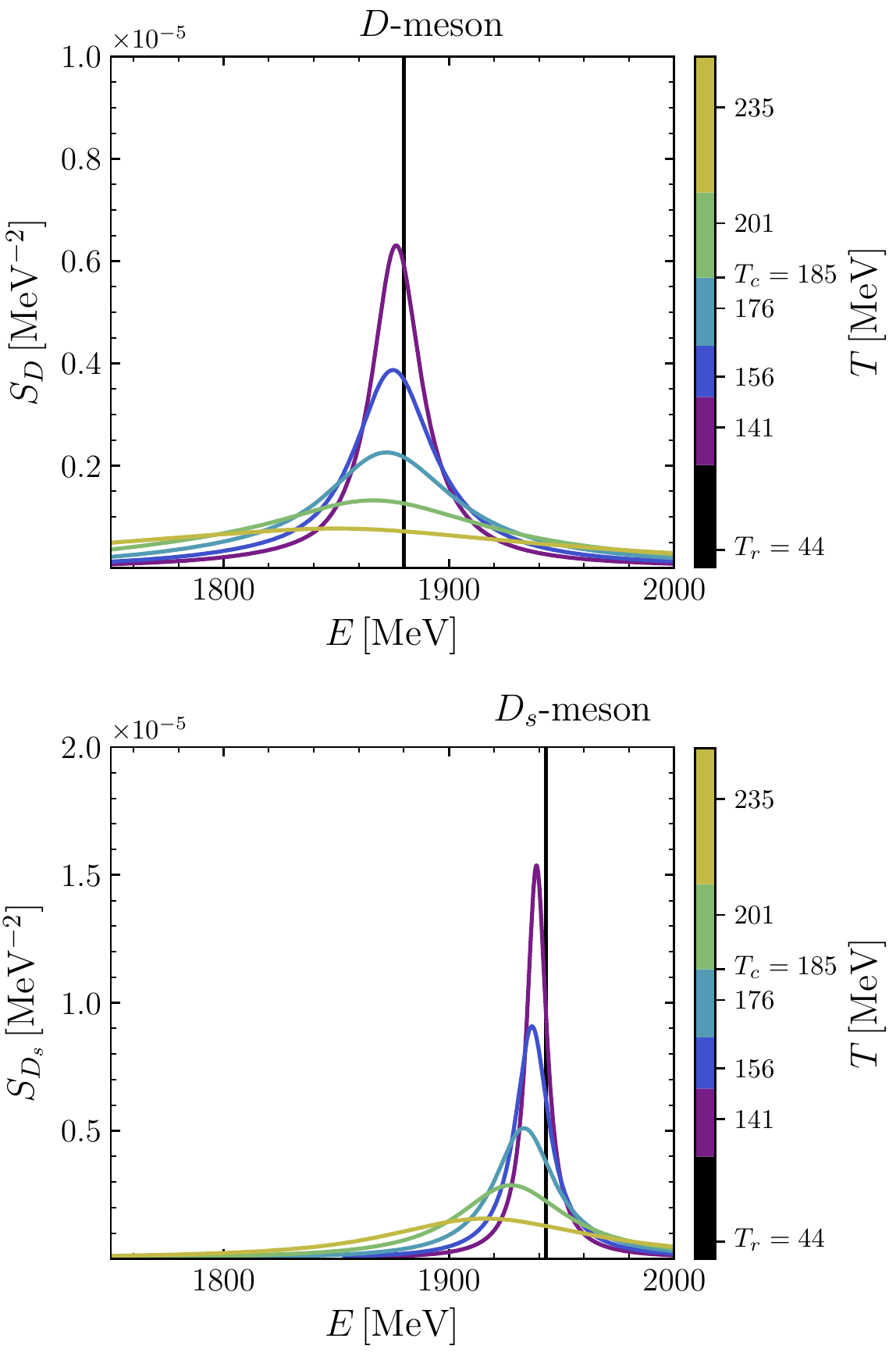}
  \caption{Spectral functions of the $D$-meson (upper panel) and $D_s$-meson (lower panel) obtained from their effective interaction with unphysically heavy pions at finite temperature.}
  \label{fig:spectral_functions}
\end{figure}

The spectral functions of the $D$ and $D_s$ mesons obtained from their interaction with the unphysical heavy pions in a hot medium are shown in Fig.~\ref{fig:spectral_functions}.  In these figures we show the spectral functions for $D$ (upper panel) and $D_s$ (lower panel) as a function of the meson energy for the different temperatures used in Ref.~\cite{Kelly:2018hsi}. As discussed in Refs.~\cite{Montana:2020lfi,Montana:2020vjg},  we clearly see the increased broadening of both spectral functions with temperature due to the larger available phase space for decay at finite temperature. Moreover, the maximum of both spectral functions slightly moves to lower energies with temperature given the attractive character of the heavy meson - light meson interaction.

A study of net charm fluctuations \cite{Bazavov:2014yba} suggests that open charm hadrons start to dissolve already close to the chiral crossover. Although the deconfined degrees of freedom necessary to investigate the melting of charm hadrons are absent in the model, we still show the spectral functions coming from our hadronic model for temperatures above and close the lattice pseudocritical temperature $T_c$ in order to explore the validity of our results for the correlators at those temperatures, as data below $T_c$ is scarce.

\section{Results of Euclidean correlators and comparison with lattice QCD}
\label{sec:euclidean-effective}

\begin{figure*}[htbp!]
\centering
\includegraphics[scale=0.68]{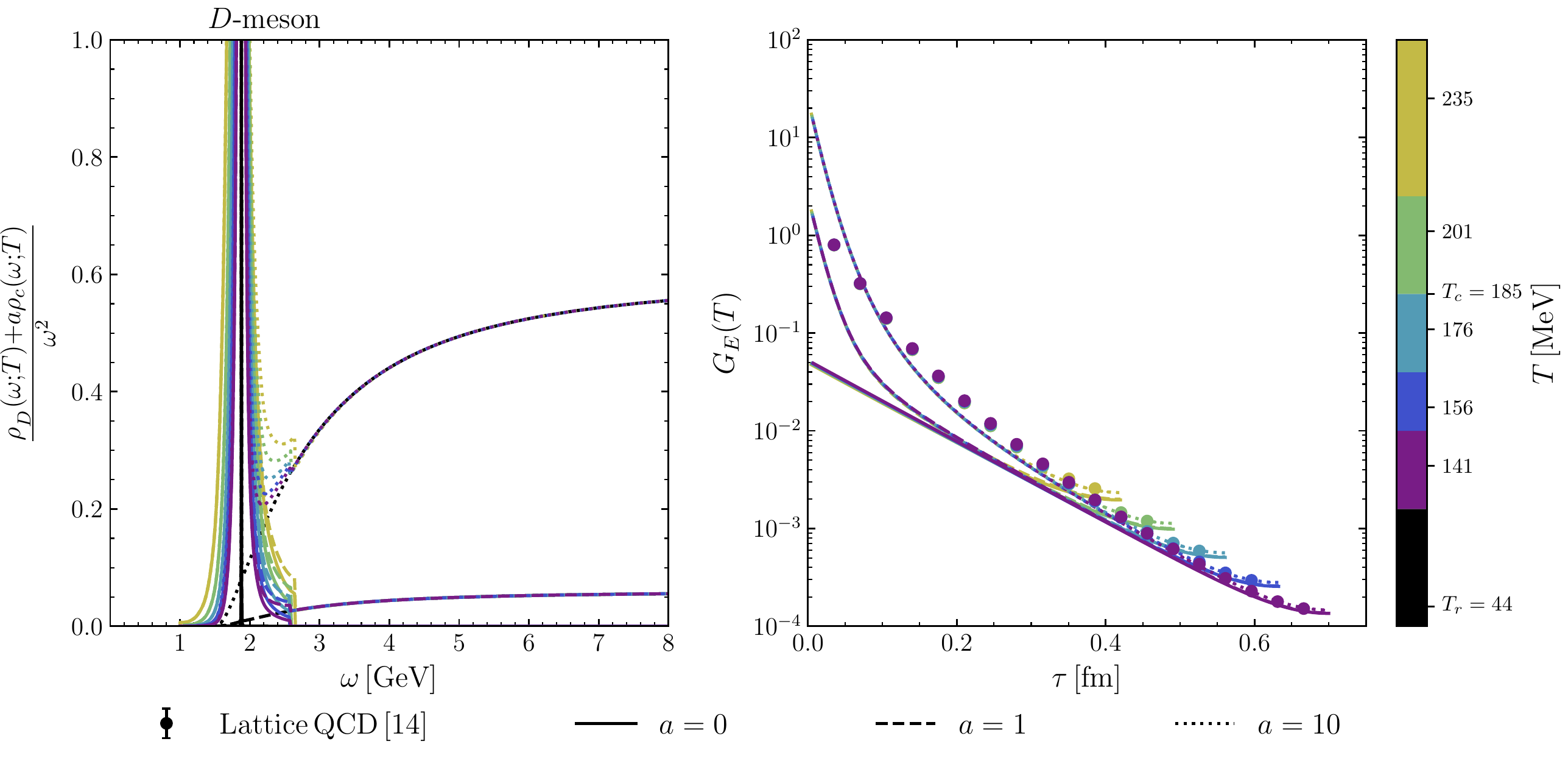}
  \caption{The $D$-meson spectral functions (left panel) and the Euclidean correlators (right panel) at different temperatures and values of the weight of the continuum spectral function (value of parameter $a$).}
  \label{fig:corr_D_unphys}
\end{figure*}

\begin{figure*}[htbp!]
\centering
\includegraphics[scale=0.68]{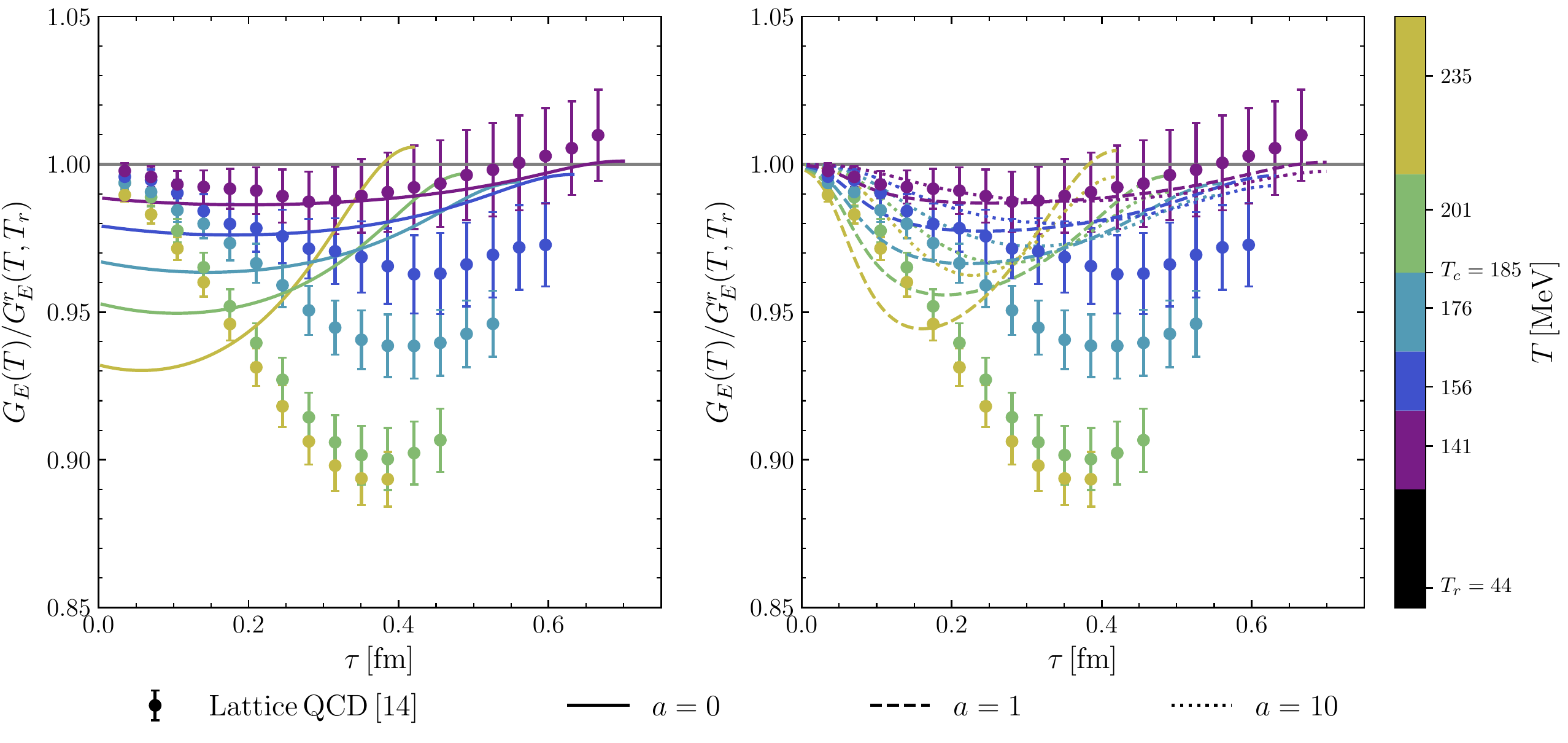}
  \caption{Ratio of the Euclidean correlators for different temperatures of Fig.~\ref{fig:corr_D_unphys}  and the reconstructed correlators at $T_r=44\,\rm MeV$, considering spectral functions with the ground state only (left panel) and adding a continuum with a weight $a$ (right panel).}
  \label{fig:ratio_D_unphys}
\end{figure*}

\begin{figure*}[htbp!]
\centering
\includegraphics[scale=0.68]{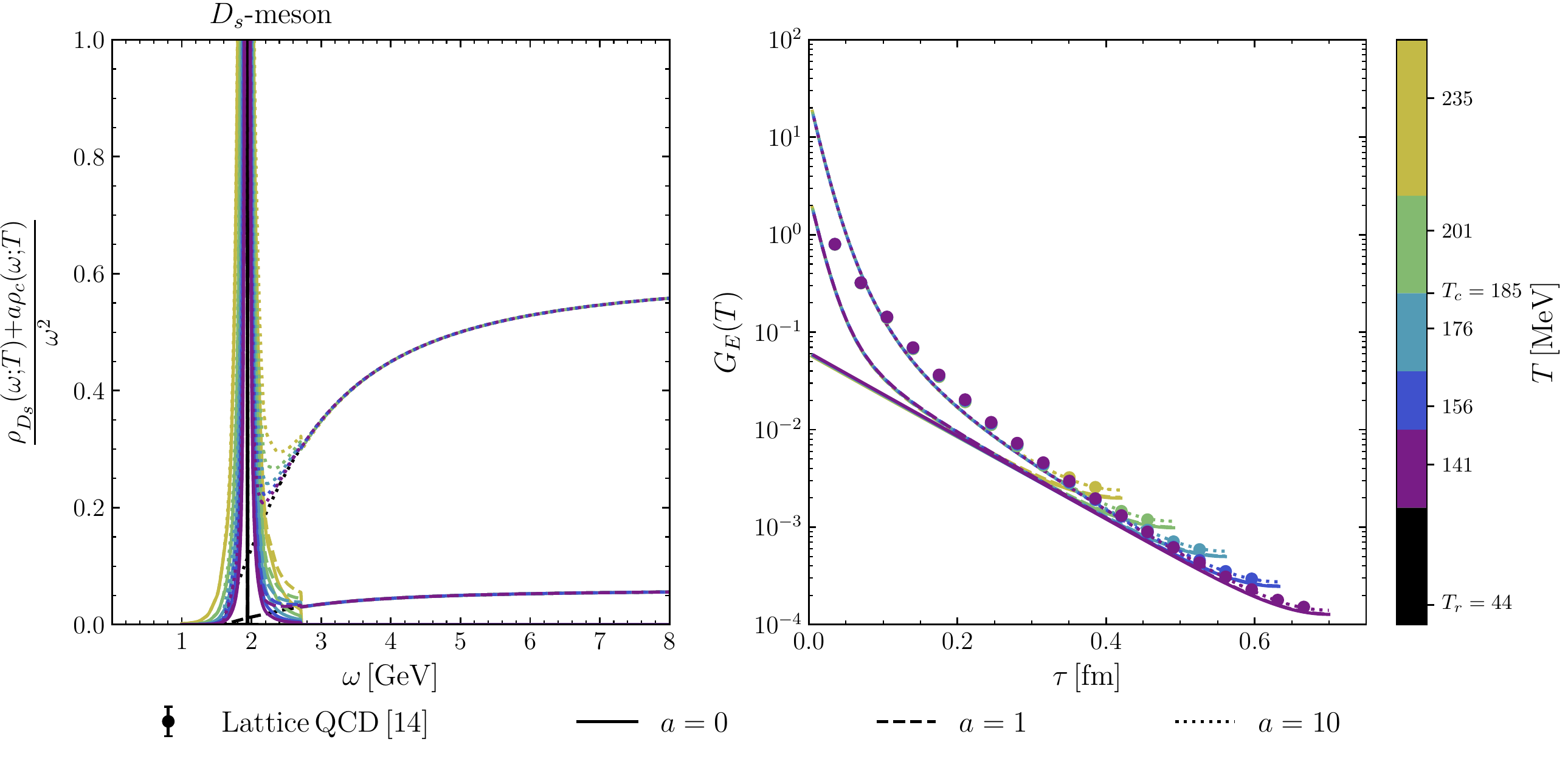}
  \caption{The $D_s$-meson spectral functions (left panel) and the Euclidean correlators (right panel) at different temperatures and values of the weight of the continuum spectral functions (value of parameter $a$).}
  \label{fig:corr_Ds_unphys}
\end{figure*}

\begin{figure*}[htbp!]
\centering
\includegraphics[scale=0.68]{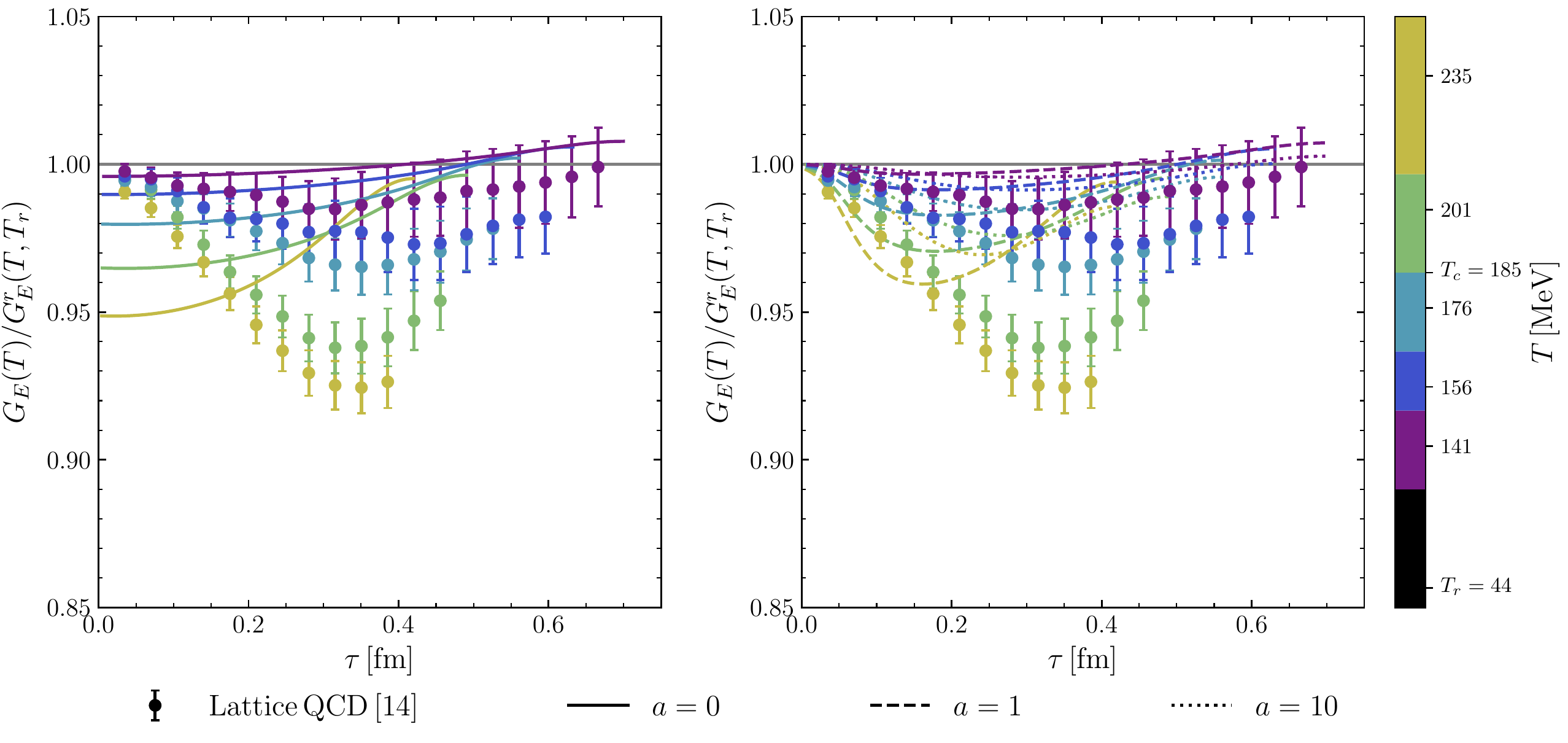}
  \caption{Ratio of the Euclidean correlators for different temperatures of Fig.~\ref{fig:corr_Ds_unphys} and the reconstructed correlators at $T_r=44\,\rm MeV$, considering spectral functions with the ground state only (left panel) and adding a continuum with a weight $a$ (right panel).}
  \label{fig:ratio_Ds_unphys}
\end{figure*}

Once the $D$ and $D_s$ spectral functions at finite temperature are known, we can obtain the corresponding Euclidean correlators from
Eqs.~(\ref{eq:corr_to_sfunc}) and (\ref{eq:corr_reconstructed}) so as to compare them to lattice QCD calculations. It is important to notice that the spectral function $S_D(\omega;T)$ of Eq.~(\ref{eq:Sfunc}), displayed in Fig.~\ref{fig:spectral_functions}, differs from $\rho(\omega;T)$, entering in Eq.~(\ref{eq:corr_to_sfunc}). This is due to the fact that the former contains the ground-state peak and an additional continuum, corresponding to $D\pi\pi$ scattering states in the case of the $D$ spectral function and $D\pi K$ states in the case of the spectral function of the $D_s$, while the latter contains all possible quark-antiquark ($c\overline{l}$ or $c\overline{s}$) states. 
 
Furthermore, the dimensions of $S_D(\omega;T)$ are MeV$^{-2}$ while the dimensions of $\rho(\omega;T)$ are MeV$^2$.  They are related through the fourth power of the charm meson mass  \cite{Klingl:1996by,Gubler:2016itj}: 
\begin{equation}
 \rho_{\rm gs}(\omega;T)=M_D^4S_D(\omega;T),
\end{equation}
with $M_D$ the vacuum value in the PDG for the mass of the $D$ (or $D_s$) meson. Note also that in lattice QCD studies the reconstructed spectral function is usually identified with the dimensionless quantity $\rho(\omega;T)/\omega^2$.

 
With $\rho_{\rm gs}(\omega;T)$ and $\rho_{\rm gs}(\omega;T_r)$ we can readily calculate Euclidean correlators and the ratios with the reconstructed correlators, $G_E(\tau;T)/G_E^r(\tau;T,T_r)$, for a direct comparison with lattice data. The input spectral function at $T=(141,\,156,\,176,\,201,\,235)$ MeV and $T_r=44$ MeV are shown with solid lines in the left panel of Fig.~\ref{fig:corr_D_unphys} for the $D$ meson and in the left panel of Fig.~\ref{fig:corr_Ds_unphys} for the $D_s$ meson, while the Euclidean correlators are displayed in solid lines in the corresponding right panels together with the lattice data of Ref.~\cite{Kelly:2018hsi} (coloured filled circles). Also, in the left panels of Figs.~\ref{fig:ratio_D_unphys} and \ref{fig:ratio_Ds_unphys} the ratios for the correlators are shown with solid lines together with the lattice results of Ref.~\cite{Kelly:2018hsi} (coloured filled circles with error bars).
 
The first clear observation is the deviation of the correlators  (solid lines in the right panels of Figs.~\ref{fig:corr_D_unphys}  and \ref{fig:corr_Ds_unphys}) and the ratio of correlators (left panels of Figs.~\ref{fig:ratio_D_unphys} and \ref{fig:ratio_Ds_unphys}) at small Euclidean times $\tau$ with respect to the lattice data points for all the temperatures and for both the $D$ and $D_s$ mesons. However, we note that for the lowest temperature $T=141$ MeV, the ratio of correlators lies within the error bars of the lattice data. For increasing temperatures, the calculated ratios deviate largely from the lattice calculations. Above or close to the pseudocritical temperature, $T_c=185$ MeV, we do not expect a good matching as the deconfined degrees of freedom are not included in the hadronic model described above, but one would expect a better comparison at lower temperatures. 

The discrepancy observed at small $\tau$ for temperatures below $T_c$ might be due to the fact that the spectral functions do not contain the higher-energy states present in the lattice correlators in addition to the ground-state, i.e. possible excited states and the continuum spectrum. As a first approximation, in the following we only add a continuum contribution to the spectral functions. In this way, we aim at understanding the differences with the fewest possible parameters, while trying to improve the comparison of the hadronic and lattice approaches.

With this goal we define the lattice spectral function as
\begin{equation}
 \rho(\omega;T)= \rho_{\rm gs}(\omega;T)+a\,\rho_{\rm cont}(\omega;T) ,
\end{equation}
where we add, to the ground-state spectral function obtained from the effective field theory, the contribution of a continuum of scattering states weighted with a factor $a$.  

The continuum contribution to the spectral function is sometimes mimicked with a step function. A parametrization of the free meson spectral function in the non-interac\-ting limit is also often used and was first derived for charmonium states in Refs.~\cite{Karsch:2000gi,Karsch:2003wy}. This spectral function describes quark-antiquark pairs with degenerate masses in the limit of infinitely high temperature. An equivalent expression can be derived in the case of non-degenerate quark masses $m_1>m_2$ \cite{FMeyerThesis}
\begin{align}
 \rho_M(\omega;T)&=\frac{N_c}{32\pi}\sqrt{\Big(\frac{m_1^2-m_2^2}{\omega^2}+1\Big)^2-\frac{4m_2^2}{\omega^2}}\,\omega^2 \\ \nonumber &\times \Big[(a_M-b_M)+2b_M\frac{m_1^2+m_2^2}{\omega^2}-4c_M\frac{m_1m_2}{\omega^2}\\ \nonumber &-(a_M+b_M)\Big(\frac{m_1^2-m_2^2}{\omega^2}\Big)^2\Big] \\ \nonumber &\times[n(-\omega_0,T)-n(\omega-\omega_0,T)]\theta\,(\omega-(m_1+m_2))\,,
\end{align}
where $N_c=3$ is the numbers of colours, $\omega_0=\frac{1}{2\omega}(\omega^2+m_1^2-m_2^2)$, $n(\omega,T)=[e^{\omega/T}+1]^{-1}$ is the Fermi-Dirac distribution and the coefficients $(a_M,b_M,c_M)$ are $(1,-1,1)$ for the scalar, $(1,-1,-1)$ for the pseudoscalar, $(2,-2,-4)$ for the vector, and $(2,-2,4)$ for the pseudovector channels. Therefore, for pseudoscalar mesons we have
\begin{align}
 \rho_{\rm cont}(\omega;T)&=\frac{3}{32\pi}\sqrt{\Big(\frac{m_1^2-m_2^2}{\omega^2}+1\Big)^2-\frac{4m_2^2}{\omega^2}}\,\omega^2 \\ \nonumber &\times 2\Big(1-\frac{(m_1-m_2)^2}{\omega^2}\Big)\\ \nonumber &\times [n(-\omega_0,T)-n(\omega-\omega_0,T)]\,\theta(\omega-(m_1+m_2))\,.
\end{align} 

In the case of the $D$ meson we take $m_1=m_c=1.5$ GeV and $m_2=m_l=0$, whereas for $D_s$  we use $m_1=m_c=1.5$ GeV and $m_2=m_s=100$ MeV. The left panels of Figs.~\ref{fig:corr_D_unphys} and \ref{fig:corr_Ds_unphys} contain the spectral functions obtained for three different values of the continuum to ground-state contribution: $a=0$ (solid lines, no continuum), $a=1$ (dashed lines) and $a=10$ (dotted lines), for the $D$ and $D_s$, respectively. The corresponding euclidean correlators are plotted in the right panels of Figs.~\ref{fig:corr_D_unphys} and \ref{fig:corr_Ds_unphys} and the ratios with the reconstructed correlators are shown in Figs.~\ref{fig:ratio_D_unphys} and \ref{fig:ratio_Ds_unphys}.

The inclusion of the continuum in the spectral functions improves the behaviour of the correlators and the ratio of correlators at small $\tau$, but does not allow to reproduce the shape of the lattice correlators\footnote{Note that the lattice correlators we used here are not continuum extrapolated and therefore suffer from cut-off effects at small $\tau \lesssim 0.1~\textrm{fm}$.}. It also permits the ratios to go to one at $\tau\rightarrow 0$ for all temperatures, as the region of very small Euclidean times is essentially governed by the spectral function at very high energies. 
However, the modification of the ratios at larger $\tau$ due to the inclusion of the continuum is rather moderate and only the results for the lowest temperature of $T=141$ MeV are compatible with the lattice data within the error bars. We have checked that this region is rather sensitive to the shape of the spectral functions at energies around $\omega=3$ GeV, where the free spectral function might be not enough to describe the continuum and where excited states are likely to be present.


\section{Conclusions and Outlook}
\label{sec:conc}

In this work we have computed for the first time Euclidean correlators for the charm $D$ and $D_s$ mesons from their corresponding thermal spectral functions obtained within a finite-temperature self-consistent unitarized approach based on a chiral effective field theory that implements heavy-quark spin symmetry. The goal is to compare the calculated correlators with those obtained in lattice QCD simulations at unphysical masses.

We have started by analyzing the behaviour of the Euclidean correlators once the full energy-dependent spectral functions are considered and have found that for temperatures well below the deconfinement transition temperature the ratio of correlators lies within the error bars of the lattice data. However, as we increase the temperature, the ratios deviate significantly from the lattice predictions due to the fact that the spectral functions do not contain the higher-energy components of the bound excited states as well as the continuum of scattering states that are present in the lattice correlators in addition to the ground-state. 

We have therefore studied the addition of a continuum contribution to the spectral functions, so as to determine its effect on the Euclidean correlators at finite temperature, while aiming at improving the comparison between the hadronic and lattice QCD calculations. The inclusion of the continuum in the spectral functions improves the behaviour of the correlators at small $\tau$ for all the temperature studied, but it is not sufficient to reproduce the shape of the lattice correlators for temperatures close or above the deconfinement transition temperature. Indeed, the region of intermediate to large $\tau$ is very sensitive to the possible existence of other excited states ignored in our model.

In the future we plan to improve the comparison with lattice QCD results for the correlators by considering finite cutoffs in our calculations. We also aim at computing the Euclidean correlators using physical masses, so as to predict the expected behaviour of the lattice QCD correlators for temperatures below the deconfinement transition, where our effective approach is fully justified.

\begin{acknowledgement}
The authors thank J.I. Skullerud and A. Rothkopf for kindly providing the lattice QCD data. G.M. and A.R.  acknowledge  support from the Spanish Ministerio de Econom\'ia y Competitividad (MINECO) under the project MDM-2014-0369 of ICCUB (Unidad de Excelencia ``Mar\'ia de Maeztu''), and, with additional European FEDER funds, under the contract FIS2017-87534-P. G.M. also acknowledges support from the FPU17/04910 Doctoral Grant from MINECO and from a STSM Grant from THOR COST Action CA15213. L.T. acknowledges support from the FPA2016-81114-P Grant from the former Ministerio de Ciencia, Innovaci\'on  y  Universidades,  the PID2019-110165GB-I00 Grant from the Ministerio de Ciencia e Innovaci\'on, the Heisenberg  Programme  of the Deutsche Forschungsgemeinschaft (DFG, German research Foundation) under the Project Nr. 383452331 and Nr. 411563442 (Hot Heavy Mesons), and the THOR COST Action CA15213.  L.T. and O.K. acknowledge support from the DFG through the Grant Nr. 315477589 - TRR 211 (Strong-interaction matter under extreme conditions).
\end{acknowledgement}

%
%


%


 \bibliographystyle{spphys}
 \bibliography{correlators}

\end{document}